# 3DPX: Progressive 2D-to-3D Oral Image Reconstruction with Hybrid MLP-CNN Networks

Xiaoshuang Li[1,2*][0000−0002−0773−6462], Mingyuan Meng[1,2*][0000−0002−9562−1613], Zimo Huang[2][0009−0005−5808−6251], Lei Bi[1,2][0000−0001−9759−0200], Eduardo Delamare[2][0000−0002−4866−1946], Dagan Feng[2][0000−0002−3381−214X], Bin Sheng[1][✉][0000−0001−8510−2556], and Jinman Kim[2][0000−0001−5960−1060]

[1] Shanghai Jiao Tong University, Shanghai 200240, China
shengbin@sjtu.edu.cn
[2] The University of Sydney, Sydney NSW 2000, Australia

**Abstract.** Panoramic X-ray (PX) is a prevalent modality in dental practice for its wide availability and low cost. However, as a 2D projection image, PX does not contain 3D anatomical information, and therefore have limited use in dental applications that can benefit from 3D information, i.e., tooth angular misalignment detection and classification. Reconstructing 3D structures directly from 2D PX has recently been explored to address limitations. However, existing methods primarily rely on Convolutional Neural Networks (CNNs) for 2D-to-3D mapping which remains a challenge to correctly infer depth-axis spatial information. In addition, these methods are limited by the intrinsic locality of convolution operations, as the convolution kernels only capture the information of immediate neighborhood pixels. In this study, we propose a progressive hybrid Multilayer Perceptron (MLP)-CNN pyramid network (3DPX) for 2D-to-3D oral PX reconstruction. Our method was motivated by the recent advancement of MLP works, which . Our 3DPX extends the MLP definition with a progressive reconstruction strategy, where 3D images are progressively reconstructed in a pyramid network with guidance imposed on the intermediate reconstruction result at each of pyramid level. Our 3DPX further captures long-range visual dependence by integrating MLPs and CNNs, thus improving semantic understanding during the reconstruction process. Extensive experiments with two large datasets involving 464 studies demonstrate that our 3DPX outperforms state-of-the-art 2D-to-3D oral reconstruction methods, including standalone MLP and transformers, in reconstruction quality, and also improves the performance of downstream angular misalignment classification tasks.

**Keywords:** Progressive Reconstruction, MLPs, Oral Panoramic X-ray.

## 1 Introduction

Panoramic X-ray (PX), an extra-oral imaging technique, is widely used in dental practices for diagnostic, assessment and monitoring purposes [1–4]. It generates stretched 2-dimensional (2D) images of the entire maxillomandibular area, enabling the evaluation of both teeth and jawbone, by rotating an X-ray emitter around the patient's head along a curved trajectory while capturing X-ray projections of the anatomical structures. When compared to other dental imaging modalities such as cone-beam computed tomography (CBCT), magnetic resonance imaging (MRI), and ultrasonography, PX has the advantages in lower cost, patient convenience, and lower radiation dose [5]. However, as a flat projection 2D image, PX lacks 3D anatomical information, which impedes accurate disease interpretation [6–8] and potential downstream tasks such as lesion segmentation, disease classification, and angular misalignment detection. Recently, there has been great research interests in reconstructing 3D structures directly from 2D oral PX [5, 9], with the aim of extracting 3D information from standard 2D PX images.

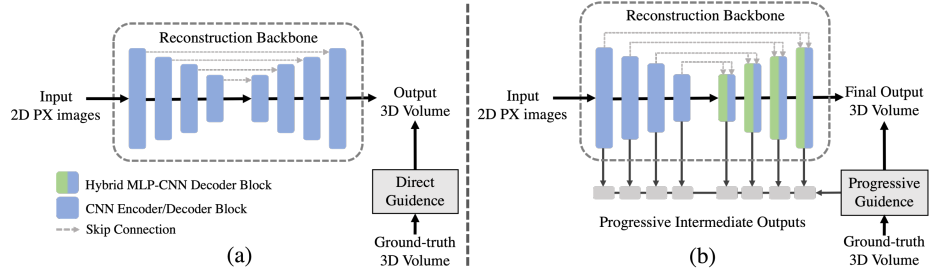

Fig. 1. A conceptual illustration of (a) regular encoder-decoder direct reconstruction architecture, and (b) our hybrid CNN-MLP progressive reconstruction strategy used in the 3DPX for 2D-to-3D panoramic X-ray reconstruction.

Existing 2D-to-3D oral PX reconstruction methods primarily rely on using Convolutional Neural Networks (CNNs) to directly predict 3D structures from 2D PX images using a 2D encoder-decoder network architecture. By using 2D convolutional layers on 3D image data, the depth information of 3D images is processed as the feature channels in the 2D convolutional layers. Specifically, at the encoder side, the network gradually reduces the image width/height and infers the depth information by successively downsampling the feature size and increasing the feature channels. At the decoder side, the network gradually recovers the width and height information with the aid of skip connections while maintaining the number of feature channels. During this process, 2D CNNs have counter-intuitive advantages over 3D CNNs, as 2D convolutional kernels leverage all depth information in the feature channels at the same time, while 3D convolutional layers only leverage part of depth information within the 3D convolution kernels such that leads to an incomplete usage of the inferred depth information. Song et al. [5] developed Oral-3D, a Generative Adversarial Network (GAN) model with a Residual CNN generator. Liang et al. [9] proposed a CNN architecture to firstly segment the PX images and then generate voxelized teeth based on the segmentation masks. Nevertheless, existing CNN-based oral PX reconstruction methods still have two key unsolved limitations:

Firstly, existing 2D-to-3D reconstruction methods employ CNNs to directly map 2D PX images to 3D image volumes (Fig. 1a). However, this simple mapping underestimates the complexity of 2D-to-3D reconstruction and cannot fully handle the difficulties in inferring depth-axis spatial information from 2D images with only height and width axes. Further, the intermediate feature maps are not fully leveraged such that the reconstruction results often lack details and tend to generate artifacts.

Secondly, CNN-based methods are limited by the intrinsic locality of convolution operations. Although 2D CNN can leverage all the depth information, its receptive fields on the height and width axes are still limited. Transformer-based models have been widely used to capture long-range dependence within images [8]. Multilayer Perceptrons (MLPs) have demonstrated strong capabilities in capturing fine-grained long-range dependence among high-resolution image details [10]. However, MLPs were designed to weighting the importance of different feature channels and improved the concentration on some channels while suppressing on others, such that has limited capabilities in preserving the coherence on the depth dimension.

In this study, we propose 3DPX, a progressive hybrid MLP-CNN pyramid network for 2D-to-3D oral PX reconstruction to overcome the two challenges mentioned in above. Our 3DPX extends the definition of MLPs with a hybrid MLP-CNN module with a 2-step skip connection to reinforce the coherence of depth information. When compared to the current state-of-the-art methods, we introduce the following contributions: (i) we introduce a progressive reconstruction strategy, where 3D images are progressively reconstructed in a pyramid network with guidance imposed on the intermediate reconstruction results at each of pyramid level (Fig. 1b), such that resulted in more fine-grained reconstruction; and (ii) Our 3DPX integrates the advantages of MLPs and CNNs, such that allows to capture long-range visual dependence and small subtle details, thus improving semantic understanding during reconstruction. To the best of our knowledge, this is the first study that introduces progressive reconstruction and MLPs for 2D-to-3D oral PX reconstruction. Extensive experiments on public (Cui et al. [11]) and our private CBCT datasets demonstrate that our 3DPX can outperform state-of-the-art 2D-to-3D oral reconstruction methods in reconstruction quality and in downstream angular misalignment classification tasks.

## 2 Method

### 2.1 Overview

We propose 3DPX to reconstruct flattened 3D structure from a single PX image, with the intention of enhancing the PX image analysis in the downstream tasks. Fig. 2 shows the workflow of 3DPX. It takes a PX image (128×256) as input and reconstructs the corresponding 3D flattened structure (128×256×128). The reconstructed results are used to perform 3D-enhanced 2D PX analysis. Our 3DPX introduced a progressive guided reconstruction strategy with hybrid MLP-CNN blocks.

3DPX is based on a customized U-Net structure. As is shown in Fig. 2, it consists of an encoder branch with four convolutional blocks, and a decoder branch with four Hybrid MLP-CNN Blocks. At the encoder side, 3DPX reduces the image width/height and infers the depth information by successively downsampling the feature size and increasing the feature channels. At the decoder side, the network gradually recovers the width and height information with the aid of skip connections while maintaining the number of feature channels. The hybrid MLP-CNN Block is applied only in the decoder branch. All decoder blocks have the same output channel size of 128, which corresponding to the depth dimension size.

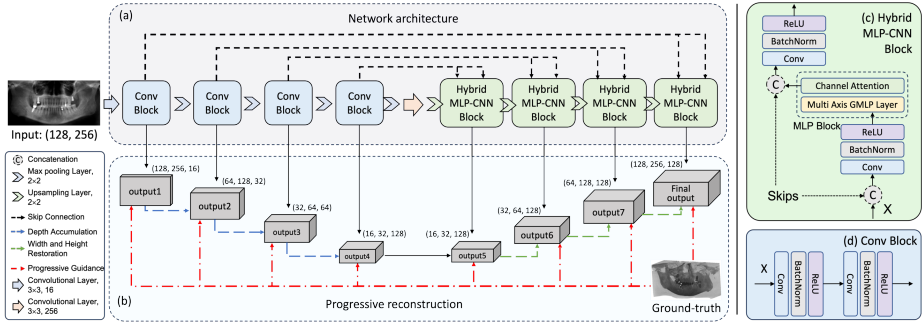

**Fig. 2.** The workflow of 3DPX for 2D-to-3D PX reconstruction. We illustrate separately the (a) network architecture and the (b) progressive reconstruction process. The core module (c) Hybrid MLP-CNN Block combines long-range attention by blending MLP and convolutional layers with a 2-step skip connection to preserve the coherence of depth information in 2D-to-3D reconstruction task.

## 2.2    Progressive Guided Reconstruction

In this section, we introduce the progressive guidance strategy for 2D-to-3D PX reconstruction and its differences from the concept of deep supervision [13, 14]. Deep supervision proposed to add auxiliary classifiers to the intermediate layers of deep neural networks to employ supervision to the shallow layers [15]. The auxiliary classifiers were optimized with the supervised task loss. It's based on the observation that in layers at different depths learn different level of features. 2D-to-3D reconstruction task presents a different scenario. Firstly, all feature maps of 3DPX are trained to imitate the 3D reconstruction successively at different scales instead of learning low-level or high-level semantic features representation. Secondly, Applying the straightforward guidance with 3D ground-truth directly on the intermediate output facilitate the feature maps into step-by-step reconstruction, while applying supervision to the auxiliary classifier improves the feature extraction at different layers.

Let $B$ be a given encoder or decoder block in 3DPX, our progressive reconstruction strategy introduces multiple guidance by applying penalties $L$ on the intermediate output of $B$. Specifically, the penalty $L = L_{SSE}(f, L)$ where $f = B(X)$ denotes the intermediate output feature map, $Y$ denotes the scaled label and $L_{SSE}$ represents the Error Sum of Square (SSE) loss. At each feature extraction stage $i$, the guided penalty is formulated as

$$L_0(X) = L_{SEE}(B_0(X), Y_0)$$
$$L_1(X) = L_{SEE}(B_1 \circ B_0(X), Y_1)$$
$$\cdots$$
$$L_i(X) = L_{SEE}(B_i \circ B_{i-1} \circ \cdots \circ B_0(X), Y_i).$$

(1)

As intermediate reconstructions progressively improve and approach the final output, the intermediate penalties should accordingly have progressive weight to emphasize the guidance close to the output layer and downplay the role of guidance close to the input layer. To achieve this, a set of hyper-parameters $\boldsymbol{\alpha}$ is set on $L_i(X)$ and the final training loss function for progressively guided reconstruction is $L_{PR}$ is formulated as

$$L_{PR} = \sum_{i=0}^{n-1} \alpha_i \cdot L_i(X).\tag{2}$$

where $n$ is the number of the encoder and decoder blocks. In the experiment, we set $\alpha_i = 2^{n-1-i}$ and for $i \leq 2$, $\alpha_i$ is empirical set to 0 to get the best reconstruction quality. Please refers to Supplementary Table. 1 for more detailed results about the ablative study of $\boldsymbol{\alpha}$.

### 2.3 3DPX with Hybrid MLP-CNN Block

The architecture of 3DPX is illustrated in Fig. 2. It's special designed U-Net structure consists of 2D Convolutional Blocks in the encoder branch and Hybrid MLP-CNN Blocks in the decoder branch. The encoder branch starts with a single convolutional layer that maps the single-channel input to a feature space with size of 16. The detail of Convolutional Block is shown in Fig. 2(d). It consists of two convolutional layers followed by Batch Normalization (BN) and ReLU activation. The output feature size in Fig. 2(a) is shown in an order of width, height, and feature channel. The encoder branch increases the size of depth channel to 128 and decrease the width and height to [16, 32]. The encoder and the decoder branch are connected by a bottleneck convolutional layer that only increase the feature channel size to 256. The Hybrid MLP-CNN Block depicted in Fig. 2(c) combines MLP Block proposed by Tu et al [16] called MAXIM. and convolutional layer together to benefit from both of their advantages. Before MLP Block, the first convolutional layer fuses the features that come from the former layer and the skip connection. In the MLP Block, the multi-axis gated MLP layer enables effective interactions between different feature spatial dimensions and capture both local and long-range dependencies of the input features. Following that, the Channel Attention mechanism in MAXIM weighting the importance of different feature channels and improved the concentration on some channels while suppressing on others. However, this emphasis treats feature channels as independent, and sabotages the coherence in the depth spatial information. The second convolutional layer do the same movement with another skip connection to integrate the long-range attention information and recover the depth coherence. The output channel size of all these layers is maintained at 128. Max pooling layers and upsampling layers with a kernel size of 2×2 are used between adjacent blocks for downsampling and upsampling.

### 2.4 Evaluation Metric

The 2D-to-3D reconstruction task was evaluated with Peak signal-to-noise ratio (PSNR), Structure similarity index (SSIM) and Dice similarity coefficient (DSC); these are standard metrics to assess image reconstruction results. DSC score was computed by extracting jaw bones from soft tissues. The bone volume was obtained by applying a threshold to the density representation. During the experiments, we set the threshold to the mean density of the ground-truth flattened 3D structure.

We further evaluated the reconstructed results via downstream application, which is on 3D-enhanced angular misalignment classification task. The classification performance was evaluated by accuracy, precision, recall, and F1 score.

## 3 Results

### 3.1 Dataset and Augmentation

We used 464 CBCT scans where 91 scans were released by Cui et al. [12], and 373 were from our private dataset. Cui et al. data were scanned in routine clinical care, where patient required dental treatments such as orthodontics, dental implants, or restoration. The images were acquired at $400 \times 400$, with a varying height of ~280 pixels, at an interslice distance of $0.4 \times 0.4 \times 0.4\ mm^3$. For our private CBCT dataset, the original resolutions were $512 \times 512$, with a varying height of ~512 pixels, at an interslice distance of $0.3 \times 0.3 \times 0.3\ mm^3$.

For PX reconstruction, dental arch curves were manually marked on all CBCTs under guidance of an experienced dentist. The synthesized PX is reconstructed by projecting the CBCT along the dental arch trajectory, in the depth direction with a 0.2 mm unit size, encompassing a depth range of 40 mm and a height of 100 mm, with the width matching the length of curve, typically around 200 mm. This region was reformatted into a flat 3D structure using curved planar reformat method. In clinical practice, the misplacement of patients' head causes angular misalignment of captured images. We augmented our dataset to acquire angular misalignment PX and corresponding flattened 3D structure by vertically and laterally rotating the CBCT scans according to [17].

Following data augmentation, we obtained 2922 pairs of PX images with dimension of [128, 256] and corresponding 3D flattened structures with dimension of [128, 256, 128]. These pairs were divided into a training set consisting of 2060 samples, an evaluation set consisting of 412 sample, and a test set consisting of 450 sample.

**Table 1.** The comparison results between the proposed 3DPX and existing 2D-to-3D reconstruction models. Best results are **bolded** and second-best results are underlined.

| Architecture | U-Net based | | | Residual CNN based | | |
|---|---|---|---|---|---|---|
| | PSNR | DSC | SSIM | PSNR | DSC | SSIM |
| CNN | 14.76 | 62.22 | 67.72 | 15.21 | 61.17 | 70.58 |
| CNN GAN (Oral-3D[5]) | 14.69 | 62.61 | 68.97 | 15.26 | 60.75 | 68.47 |
| Hybrid MLP-CNN | 14.99 | 62.2 | 68.55 | 15.42 | <u>61.64</u> | <u>72.25</u> |
| Hybrid MLP-CNN GAN | 15.11 | <u>63.7</u> | 68.42 | 15.23 | 60.96 | 69.02 |
| Progressive Hybrid MLP-CNN GAN | <u>15.51</u> | 63.21 | <u>72.17</u> | <u>15.45</u> | 60.69 | 71.22 |
| Progressive Hybrid MLP-CNN (3DPX) | **15.84** | **63.72** | **74.09** | **15.73** | **62.01** | **73.45** |

## 3.2    Experiments

The proposed 3DPX was compared with existing 2D-to-3D reconstruction models and the reconstruction results are shown in Table 1. Pure CNN-based customized U-Net achieved a SSIM of 67.72 and DSC of 62.22 on bone segmentation. Compared to U-Net, Residual CNN excels in retaining the vanished information throughout each convolution block due to its residual connection and achieved a higher SSIM of 70.58. The integration of GAN training strategy, hybrid MLP-CNN blocks and progressive guidance improved the reconstruction results. Specifically, GAN strategy elevated the SSIM to 68.97 for U-Net but decreased it to 68.47 for Residual CNN in Oral-3D. The introduction of Hybrid MLP-CNN blocks enhanced both basic models, notably increasing the SSIM for Residual CNN to 72.25 and the DSC to 61.64. Hybrid MLP-CNN introduces both the expansion of the receptive field and a detrimental effect on depth coherence. Without progressive guidance, only 2-step skip connection provide depth restoration for U-Net. But for Residual CNN, both skip and residual connection help to maintain depth coherence. However, GAN strategy failed to deliver improvement with the presence of hybrid MLP-CNN blocks, both with and without the progressive guidance. With progressive intermediate guidance, 3DPX significantly outperformed the U-Net by over 6 points in terms of SSIM, 7 points in terms of DSC, and 1.1 points on PSNR. In contrast, we only observed a moderate improvement with Residual CNN, where SSIM was increased by 2 points. When progressive guidance sufficiently restored intermediate depth coherence, the residual connection lost its efficacy and failed to surpass the more straightforward designed counterpart.

**Table 2.** Ablative study of MLP and Hybrid CNN-MLP blocks. Best results are **bolded** and second-best results are underlined.

| Method | PSNR | DSC | SSIM |
|---|---|---|---|
| UNet | 14.76 | 62.22 | 67.72 |
| 3D Decoder UNet | 14.63 | **64.32** | 62.82 |
| MLP UNet | 14.67 | 59.72 | 60.09 |
| Transformer (UNETR) | 14.76 | 60.57 | 60.3 |
| Hybrid CNN-MLP U-Net | <u>15.51</u> | 63.21 | <u>71.21</u> |
| Progressive Hybrid Unet (3DPX) | **15.84** | <u>63.72</u> | **74.09** |

Table 2 shows the ablation results. It shows that the proposed hybrid MLP-CNN block is more effective when compared to standard MLP or transformer mechanism. The results of 3D decoder U-net show a decrease when compared to 2D U-Net on structure similarity and peak signal to noise ratio. As 3D decoder focuses on local area within the 3D convolutional kernel, local continuity of high uptake area was improved according to DSC at 64.32. MLP and UNETR deteriorated the spatial continuity of reconstructions on depth channel, resulting in a decrease of 5-7 points in DSC and more than 7 points in SSIM. Hybrid CNN-MLP blocks restored and further improved the reconstruction quality based on the MLP blocks to a SSIM of 71.21 and a DSC of 63.21. With the progressive intermediate guidance, our 3DPX achieved highest score on all three metrics.

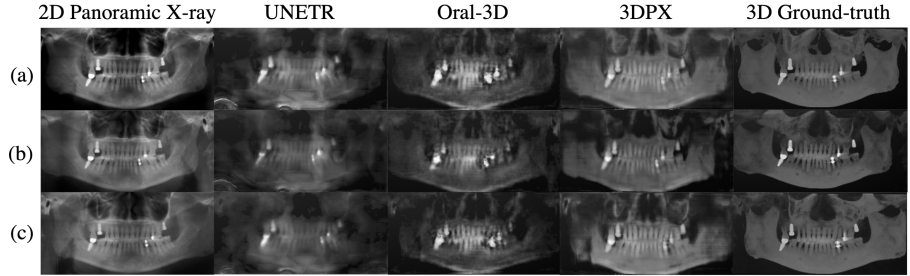

Fig. 3. Visualizations of reconstructed 3D flattened structure from PX images from comparison methods. Two types of angular misalignment augmentation are depicted, (a) regular PX capturing angle, (b) PX with left rotation misalignment, and (c) PX with right rotation misalignment.

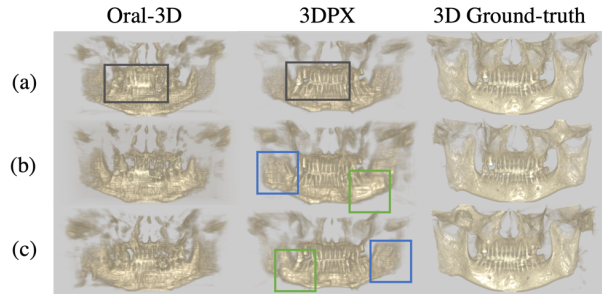

Fig. 4. Volume rendering of reconstructed 3D flattened structure of Oral-3D, 3DPX and the ground-truth.

We further conducted downstream experiments on angular misalignment classification task for PX images. The comparison happens between a 2D ResNet and joint 2D-3D ResNet designed for pure PX and 3D-enhanced PX classification. Improvement on accuracy and other metrics is observed on both binary and 5-category angular misalignment classification. Please refers to Supplementary Table. 2 for more detailed results about the downstream classification task results.

Fig. 3 and 4 provide visualizations of the reconstructed 3D flattened structure from the comparison methods alongside their volume renderings. The regions highlighted in the black box demonstrate that 3DPX generated relatively fine-grained details for the anterior teeth and tooth implants. Additionally, it produced clearer expansions of one side of the jawbone ramus in the blue box and the other side of the jawbone body in the green box, which were caused by rotation misalignment during PX capturing.

## Conclusion

In this study, we propose a progressive hybrid Multilayer Perceptron (MLP)-CNN pyramid network (3DPX) for 2D-to-3D oral PX reconstruction. Extensive experiments with two large datasets involving 464 studies demonstrate that our 3DPX outperforms state-of-the-art 2D-to-3D oral reconstruction methods, including standalone MLP and transformers, in reconstruction quality, and also improves the performance of down-stream angular misalignment classification tasks.